\documentstyle[12pt,epsfig]{article}
\RequirePackage{tabularx,float,afterpage,pifont}
\RequirePackage{rotating}

\begin{document}

\title{\Large {\bf The Micro Slit Gas Detector}} 

\vspace{1cm}



\hspace{8cm} September 18, 1998

\vspace{1cm}

\begin{center}

{\Large {\bf The Micro Slit Gas Detector}}

\vspace{.5cm}

{\small J. Claude Labb\'e}

{\small {\it EP/CMT Division CERN}}

{\small F. G\'omez\footnote{Corresponding author:\\
Faustino G\'omez\\ Departamento
de F\'{\i}sica de Part\'{\i}culas, Facultad de F\'{\i}sica\\ Universidad de
Santiago, E-15706-Santiago (Spain)\\
E--mail: Fgomez@cern.ch}, T. N\'u\~nez, A. Pazos, P. V\'azquez}

{\small {\it Universidade de Santiago de Compostela~\footnote{Work supported by Comisi\'on Interministerial de Ciencia y Tecnolog\'{\i}a (CICYT), projects AEN96-1671 and AEN97-1706-E.}}}

\end{center}

\vspace{1cm}

\begin{center}
{\large Abstract}\\
\vspace{.5cm}
\end{center}

We describe the first tests with a new proportional gas detector. Its geometry consists in slits opened in a copper metallized kapton foil with 30 $\mu$m anode strips suspended in these openings. In this way the multiplication process is similar to a standard MSGC. The fundamental difference is the absence of an insulating substrate around the anode. 
Also the material budget is significantly reduced, and the problems related to charging-up or polarization are removed. Ageing properties of this detector are under study.

\begin{center}
PACS: 2940, 2970
\end{center}

\newpage

\section{Introduction}

A new generation of high rate proportional gaseous detectors based on advanced printed circuit technology (PCB) has been introduced during the last year. Important efforts in the research and development of these kind of detectors are justified because of their low cost and robustness.  
Examples of these detectors are the Gas Electron Multiplier (GEM)~\cite{FS1}, the Micro-Groove Detector (MGD)~\cite{RB1}, and the WELL detector~\cite{RB2}. They have in common the use of thin kapton foils and PCB techniques in order to implement the multiplication structure. The flexibility of the readout is precisely another advantage of these detectors, allowing in some cases an intrinsic two dimensional device. 
Detector charging up and operation stability are important issues that need to be studied. We present here indications of a good performance for the Micro Slit Gas Detector (MSGD).

\section{Detector description}

The development of kapton etching techniques (commonly used for GEM production) has made possible the easy construction of new detector geometries.

In this case one of the metallic layers, of a 50 $\mu$m thick kapton foil copper clad on both sides, is litographically etched with a matrix of rectangular round-corner slits, 105 $\mu$m wide and 6 mm long (repeated in the transverse direction with a period of 200 $\mu$m). 
In the opposite side, a pattern 30 $\mu$m wide strips with 200 $\mu$m pitch is etched, ensuring that the strips run along the slits 
(see Figure 1).

When kapton is removed, the final device has 30 $\mu$m strips suspended only by 200 $\mu$m kapton joints regularly spaced at 8 mm (to provide mechanical stiffness) (Figure 2).
In this way a ``substrate-free'' MSGC is achieved, and the detector resembles a wire-chamber. 

The first detector prototype, 10$\times$10 cm$^2$, was enclosed in a gas volume, which was sealed symmetrically by two thin conductive foils, at 3 mm distance from the kapton plane (see Figure 3). The first provides the drift field towards the multiplication region (drift plane), and the second was given, in the test, a certain potencial with respect to the anodes, which we discuss later. Initially
this backplane was metallized with the aim to define better the electric 
field around the anode.

\section{Detector performance}

The signal development takes place in a similar way as in a standard MSGC. Drifting electrons reach the E-field region between anode and cathode, and are then multiplied inside the rectangular slits. The electron avalanche produced in this region is collected by the anode strips. The ion charge is collected by the cathode and the drifting plane, in a proportion depending on the operating voltages. 
In this case anodes were grounded through a bias resistor while a negative potential was applied to the cathode.

The detector was irradiated with X rays coming from an Cr X-ray tube and the gas mixture used was composed by Ar and DME in different proportions.

The signal was extracted from an OR of 32 anodes and amplified by a ORTEC 142PC preamplifier followed by an AFT Research Amplifier Model 2025. The output was digitized in a Tektronix TDS 684A Oscilloscope.

\subsection{Operation voltages}

Typical operating voltages in the first prototype are very similar to MSGCs. Detector gains obtained are somewhat lower\footnote{In a typical MSGC 
with 10 $\mu$m anodes and 100 $\mu$m cathodes with Ar-DME 50\% a gain
of aprox. 1000 is achieved with a cathode potential of 550, while in the
Micro Slit detector gain is around 600.}
. This is understandable due to the width of the anode strips (still limited by the PCB production technique), and also as a consequence of the extended gap between anode and cathode because of the non planar geometry and the cathode width~\footnote{New prototypes are under development with wider cathodes}. 
The detector gain exhibits an exponential dependence on the voltage applied to the cathode (Figure 4). In this Figure the maximum gains showed
were limited by sparks in the chamber. In some tests afterwards the MSGD was exposed to severe 
sparking during hours but no damage in its structure was found.

A pulse height spectrum can be seen in Figure 5. The voltage applied to the backplane does not affect essentially the anode signal, as illustrated also in Figure 5.  

Figure 6 shows spectra obtained with different values of the cathode voltage. Decreasing it by 10 V, for the drift voltage V$_{d}$=-1600 V, produces a 20$\%$ drop in the gain.   

In these spectra, the Argon scape peak is clearly separated from that corresponding to the K$_{\alpha}$ photon energy at 5.4 KeV. The energy resolution for pulse height spectra measured with V$_d$=-1500 V 
and V$_{cat}$=-515 V
is 16$\%$ FWHM, and in this field configuration 90\% of the ions 
drift to the cathode electrode.

The dependence of the gain with the cathode voltage was also studied for different gas mixtures. The results of these studies (Figure 4) show that the highest gains were obtained with high argon content in the gas mixture.  

Also the dependence on gain with the drift field is showed in Figure 7. 
Clearly an enhancement of gain is obtained with higher drift field values.

\subsection{Short term gain variation}

Typically variations on the gain during the first operation moments
manifest in those detectors using insulating substrates. This is due to the accumulation of charge on the dielectric (charging-up) and polarization, producing electric field modifications, and thus affecting the amplification process.  
Normally this effect has been avoided using higher conductive coatings (like LPVD diamond)~\cite{AB1} or substrates (like S8900).
In the GEM, for example, kapton surface of the holes is clearly traversed by the dipole electric field thus producing some charging up~\footnote{
A small admixture of water in the gas as well as straighter holes
 have demonstrated to solve the
problem.
}. In this geometry we have designed the electrodes in such a way that exposed area to E-field represents only around 1$\%$ of the total. This (see below) represents a major improvement in this type of devices just simplifying the production (no coating needed).

The effect of charging up on the MSGD gain was determined by registering the pulse height spectrum and comparing the maxima from consecutive periods. 
Figure 8 shows the evolution of the gain during the first 82 minutes of irradiation under a rate of 10$^3$ Hz mm$^2$ beginning from a cold
start (detector and beam initially switched off). Variations of the gain are less than a 4$\%$. 

In order to accelerate the effect of this possible charge accumulation, the MSGD was irradiated with a photon rate of $\approx$ 10$^{6}$ during about 10 minutes. Figure 9 compares the spectra before and after the high irradiation. No appreciable change occurs. This behaviour differs from that observed in detectors with dielectric substrate, like standard MSGC or GEM. 

\section{Rate capability}

The rate capability of this detector was determined by measuring the current in the group of instrumented anodes for different values of the incident photon flux. Driving the X-ray tube to its maximum current, we could reach up to 2.6 $\tiny{\times}$10$^6$ Hz mm$^{-2}$ incident photon flux, collimated over a surface of 3  mm$^{2}$. No appreciable drop in gain was observed. In Figure 10
the relative changes in the detector gain during the irradiation test are
shown. They were determined from the observed deviations with respect to a linear fit between X-ray intensity and anode current.

The advantage of the MSGD is the effective absence of any dielectric surface,
avoiding the use of delicate high resistive coatings to reach values of 
10$^{6}$ mm$^{-2}$ s$^{-1}$. 

\section{Conclusions}

A prototype of a new proportional gas detector, based on the PCB technology, has been designed and tested.

 The first tests with this detector show important properties, related mainly to its high rate capability (up to 2.5 MHz mm$^{-2}$) and the absence of charging up effects. 

In spite of its similarity to the MSGC in the amplification process the use of the PCB technology reduces considerably the cost and material budget. Besides, it is important to remark the supression of the substrate for supporting the anode structure.

Another interesting possibility is to set up a similar detector with a mirror cathode 
structure respect to the anode plane, thus having upper and lower drift
regions and allowing to reduce the effective drift gap and charge 
collection time. 

\section{Acknowledgements}

This work was only possible due to the invaluable help and collaboration of M. S\'anchez (CERN EP/PES/LT Bureau d' etudes Electroniques) under the
responsability of Alain Monfort, L. Mastrostefano and D. Berthet (CERN EST/SM
Section des Techniques Photom\'ecaniques).

We also thank B. Adeva, Director of the Laboratory of Particle Physics in Santiago de Compostela, where part of this work has been carried out, for his strong support and careful reading of the manuscript.

We would like to thank A. Gandi, responsible of the Section des Techniques
Photom\'ecaniques, anbd A. Placci, responsible of the Technical 
Assistance Groupe TA1, for their encouragement and logistic support.

\newpage

{\large {\bf Figure Captions}}

\vspace{.5cm}

\noindent Figure 1: Copper clad kapton design
of the Micro Slit Gas Detector (top view).

\noindent Figure 2: Scheme of one slit (transverse section).
The copper layer is 15 $\mu$m thick.

\noindent Figure 3: Schematic view of the tested prototype.

\noindent Figure 4: Behaviour of the gain as a function of the 
cathode voltage for different gas mixtures.

\noindent Figure 5: Pulse height spectra obtained with different
values of the voltage in the backplane.

\noindent Figure 6: Effect of the cathode voltage in the response of
the detector.

\noindent Figure 7: Gain dependence with the drift field.

\noindent Figure 8: Evolution of the gain during the first 
irradiation moments.

\noindent Figure 9: Pulse height spectra before and after high 
irradiation.

\noindent Figure 10: Rate capability of the Micro Slit Gas Detector.

\newpage

\pagestyle{empty}

\voffset=0cm

\begin{figure}
\begin{center}

\vspace{2cm}

\begin{turn}{0}
\mbox{\epsfig{file=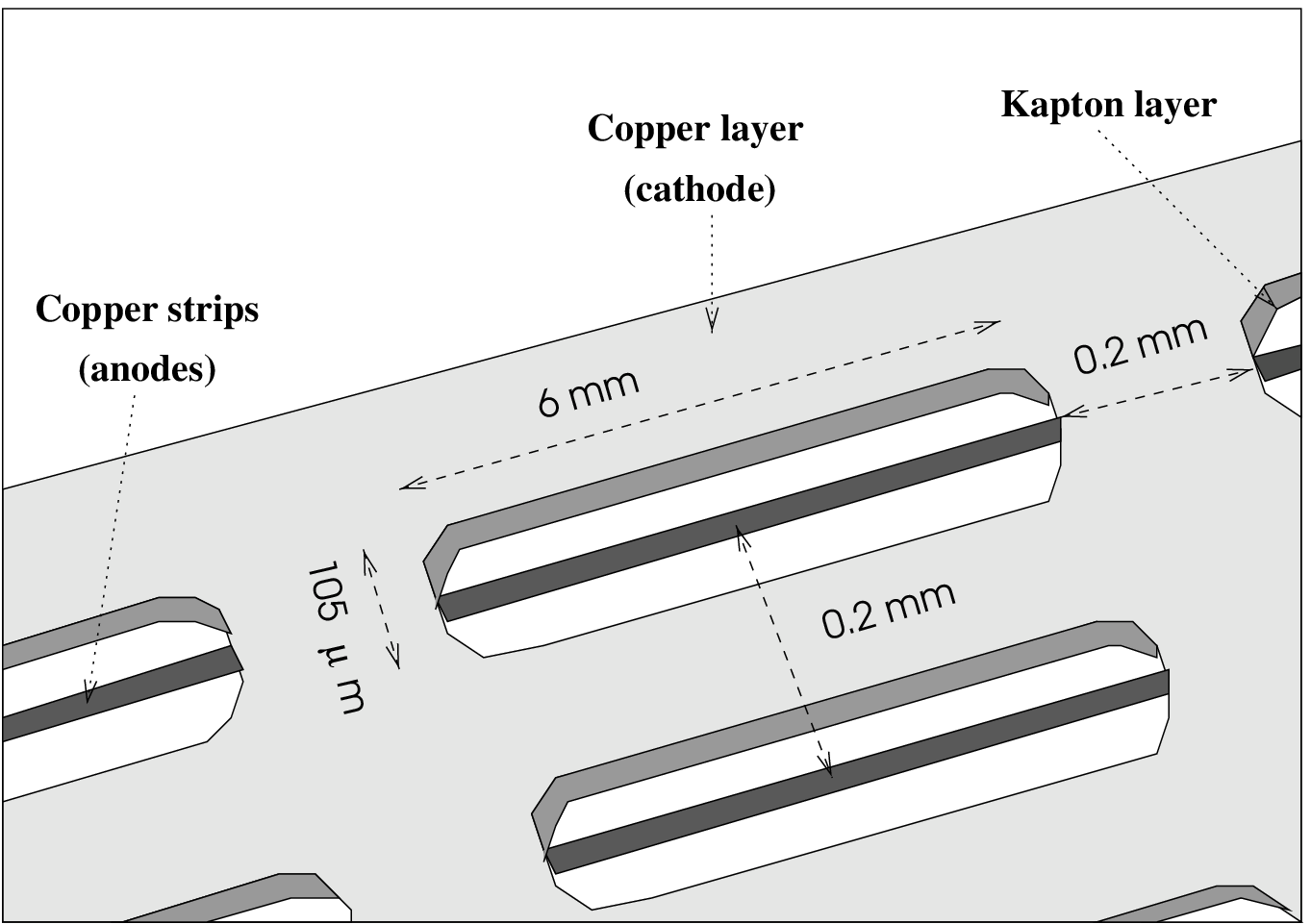,height=7.5cm}}
\end{turn}

\vspace{.5cm}

Figure 1

\vspace{5cm}

\begin{turn}{0}
\mbox{\epsfig{file=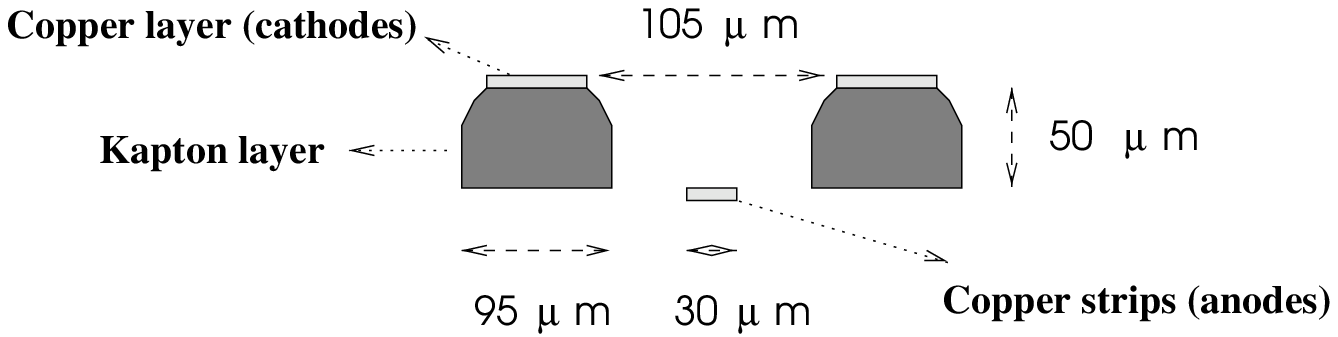,height=3.cm}}
\end{turn}

\vspace{.5cm}

Figure 2
\end{center}
\end{figure}

\newpage

\vspace{3cm}

\begin{figure}
\begin{center}
\mbox{\epsfig{file=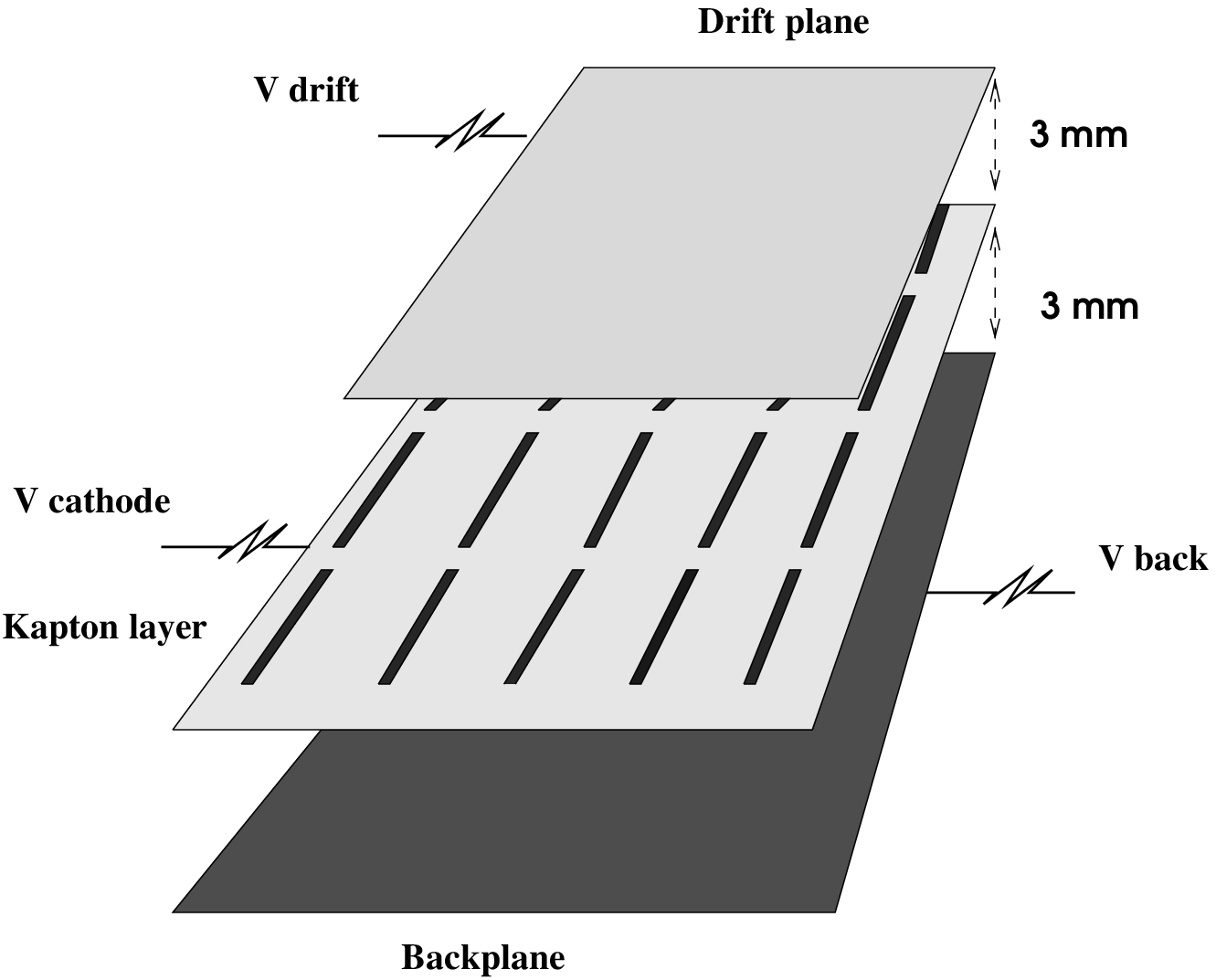,height=10cm}}

\vspace{.5cm}

Figure 3
\end{center}
\end{figure}

\newpage



\vspace{3cm}

\begin{figure}
\begin{center}
\mbox{\epsfig{file=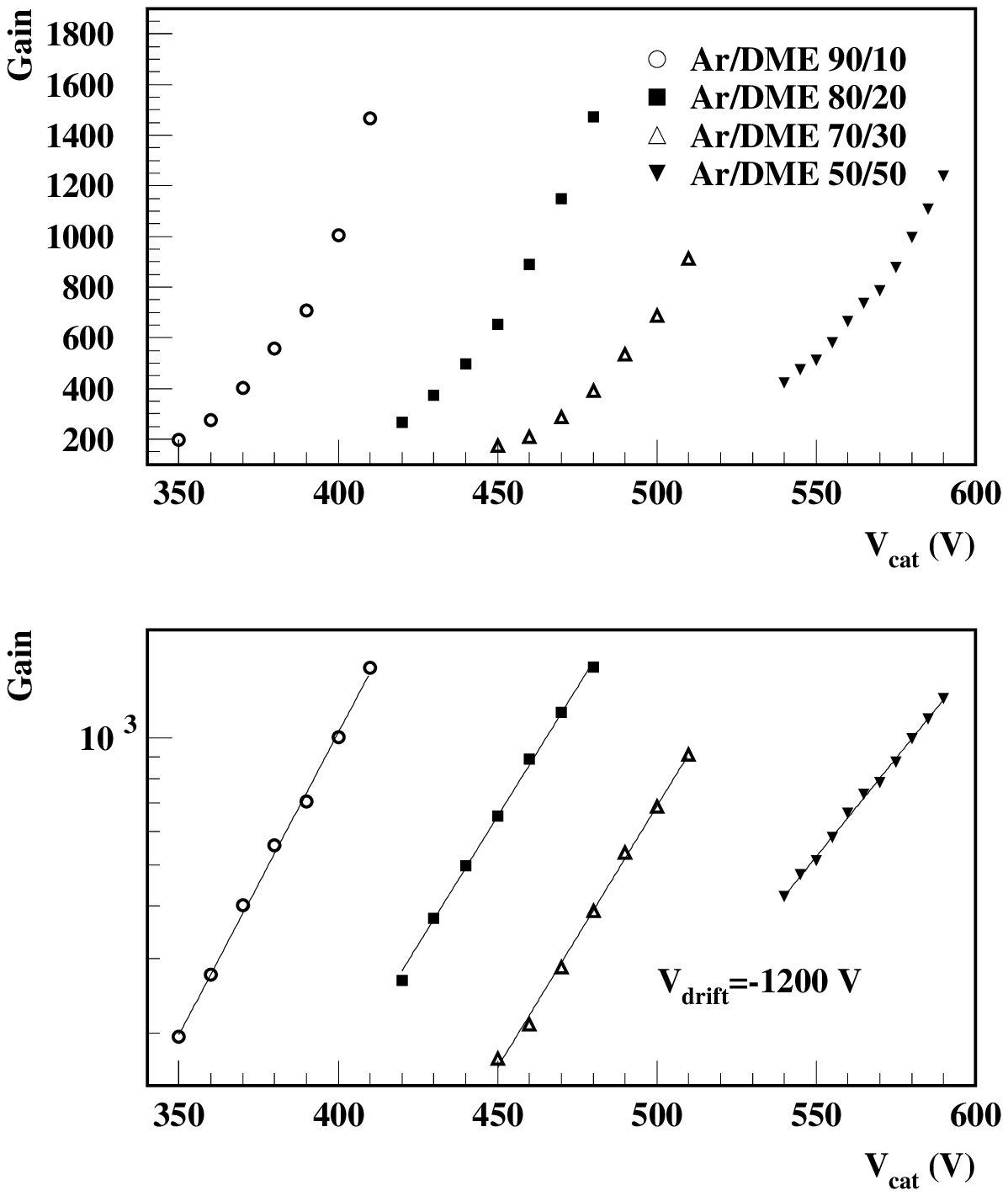,height=18cm}}

\vspace{.5cm}

Figure 4
\end{center}
\end{figure}

\newpage

\voffset=-3cm
\begin{figure}
\begin{center}
\mbox{\epsfig{file=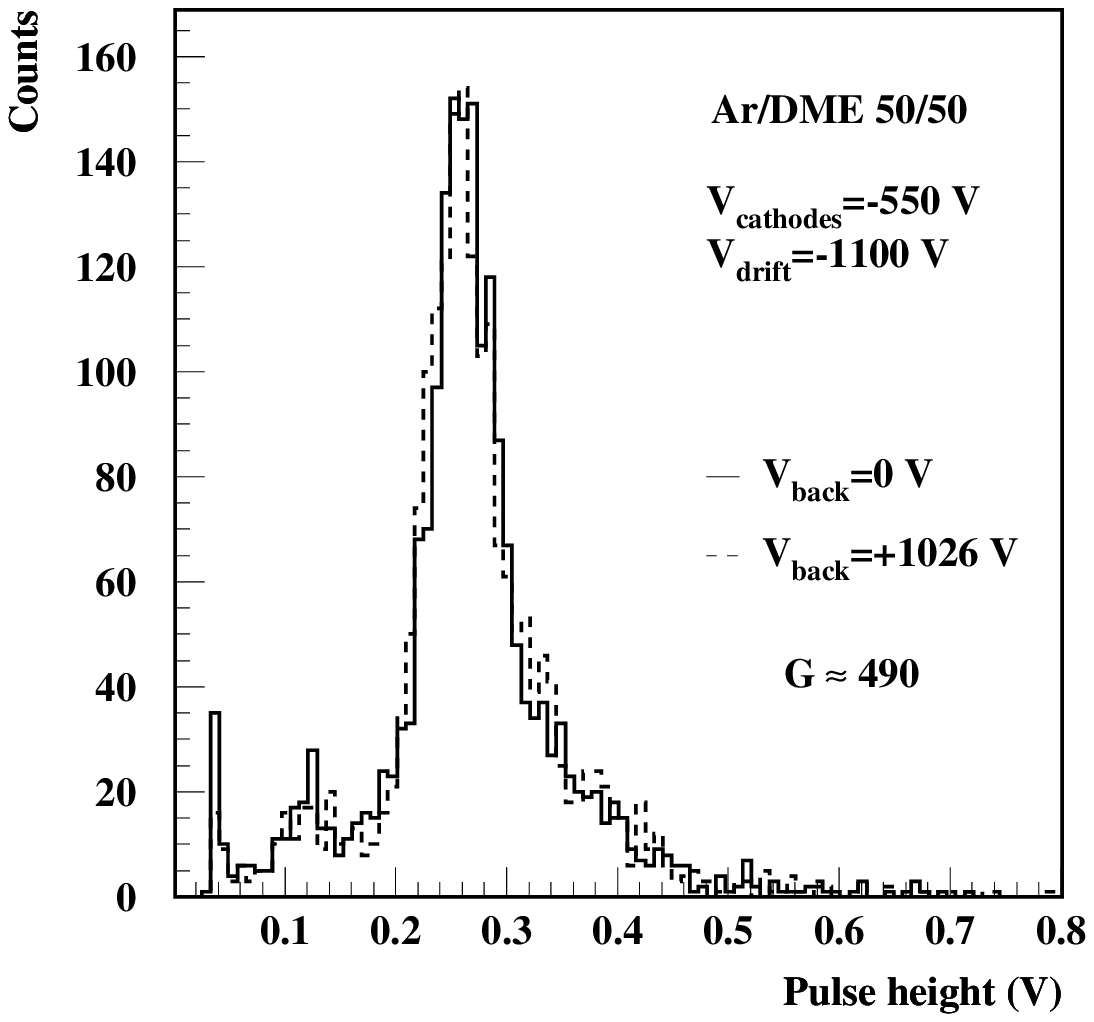,height=7.5cm,width=10cm}}

Figure 5

\mbox{\epsfig{file=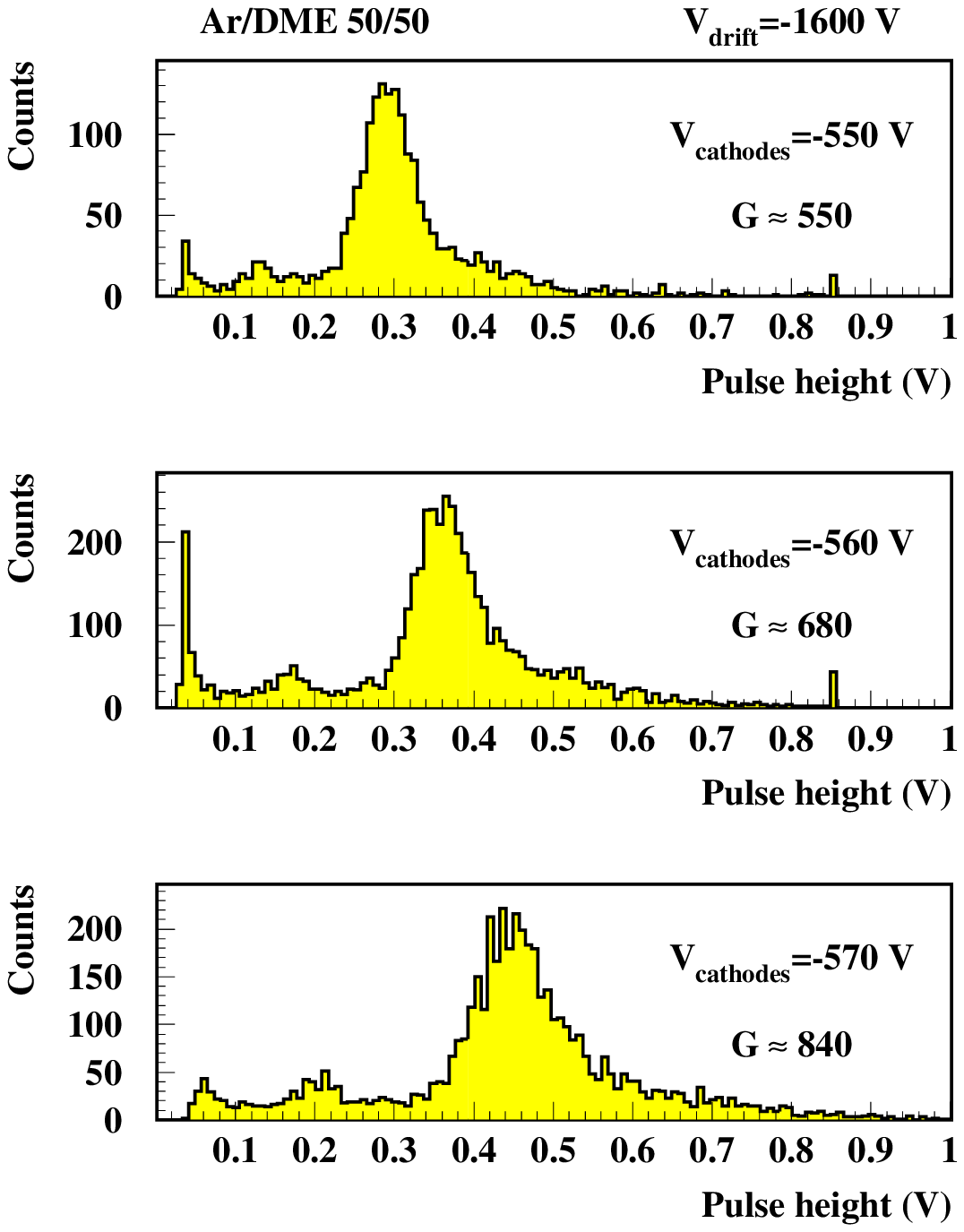,height=15cm}}

Figure 6
\end{center}
\end{figure}

\newpage

\begin{figure}
\begin{center}
\mbox{\epsfig{file=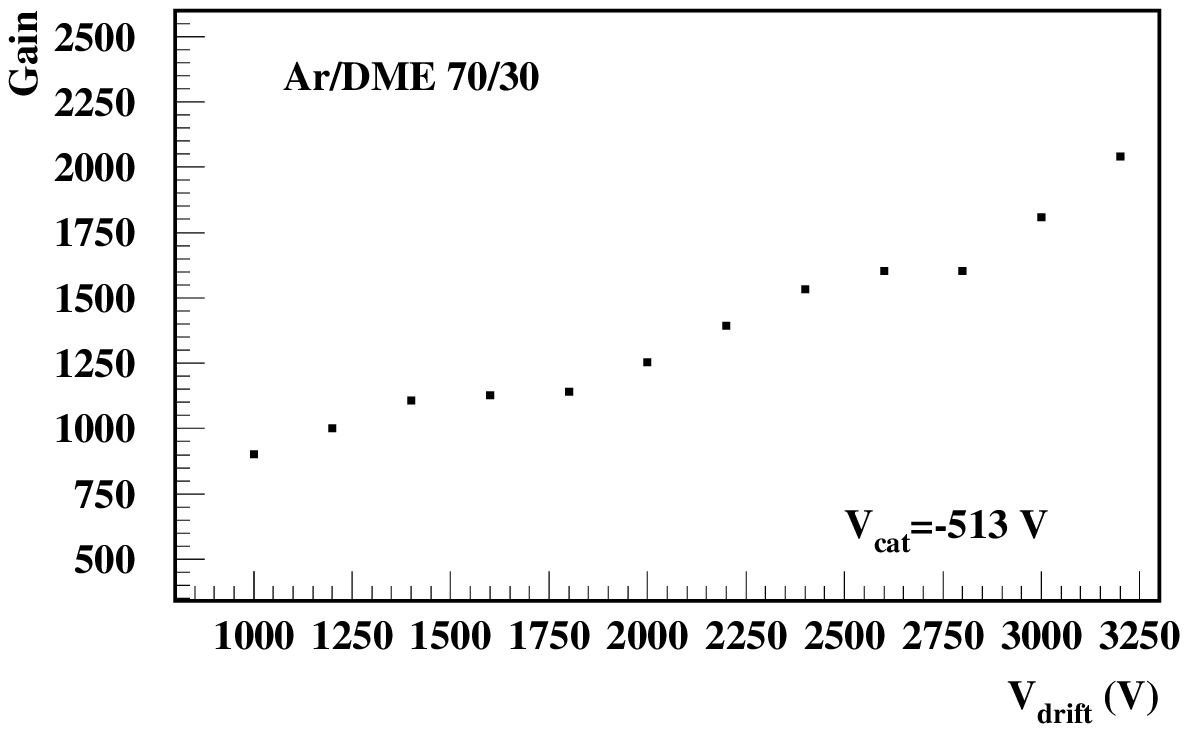,height=11cm}}

\vspace{.5cm}

Figure 7

\mbox{\epsfig{file=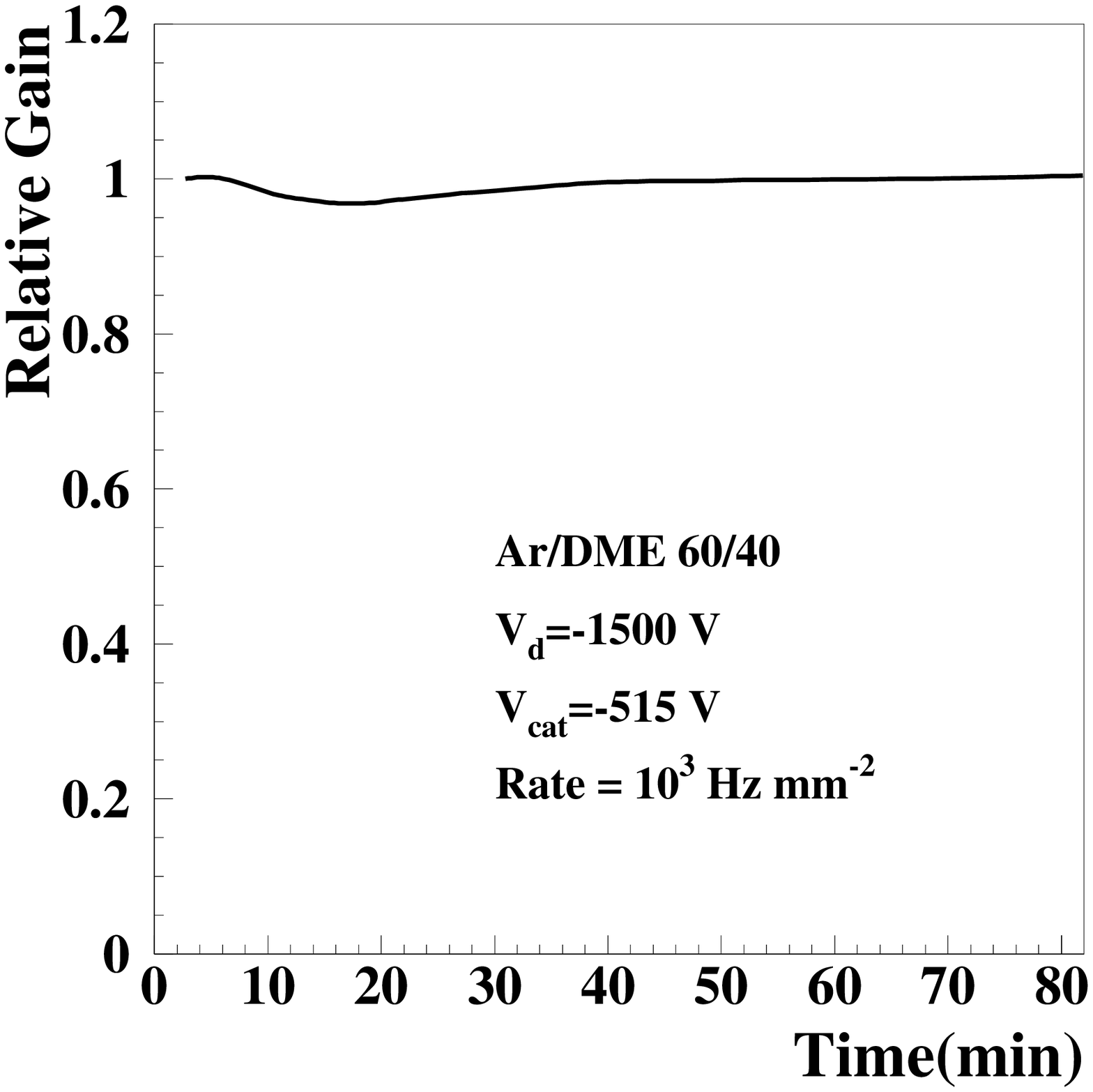,height=13cm}}

\vspace{.5cm}

Figure 8
\end{center}
\end{figure}

\newpage

\begin{figure}
\begin{center}
\mbox{\epsfig{file=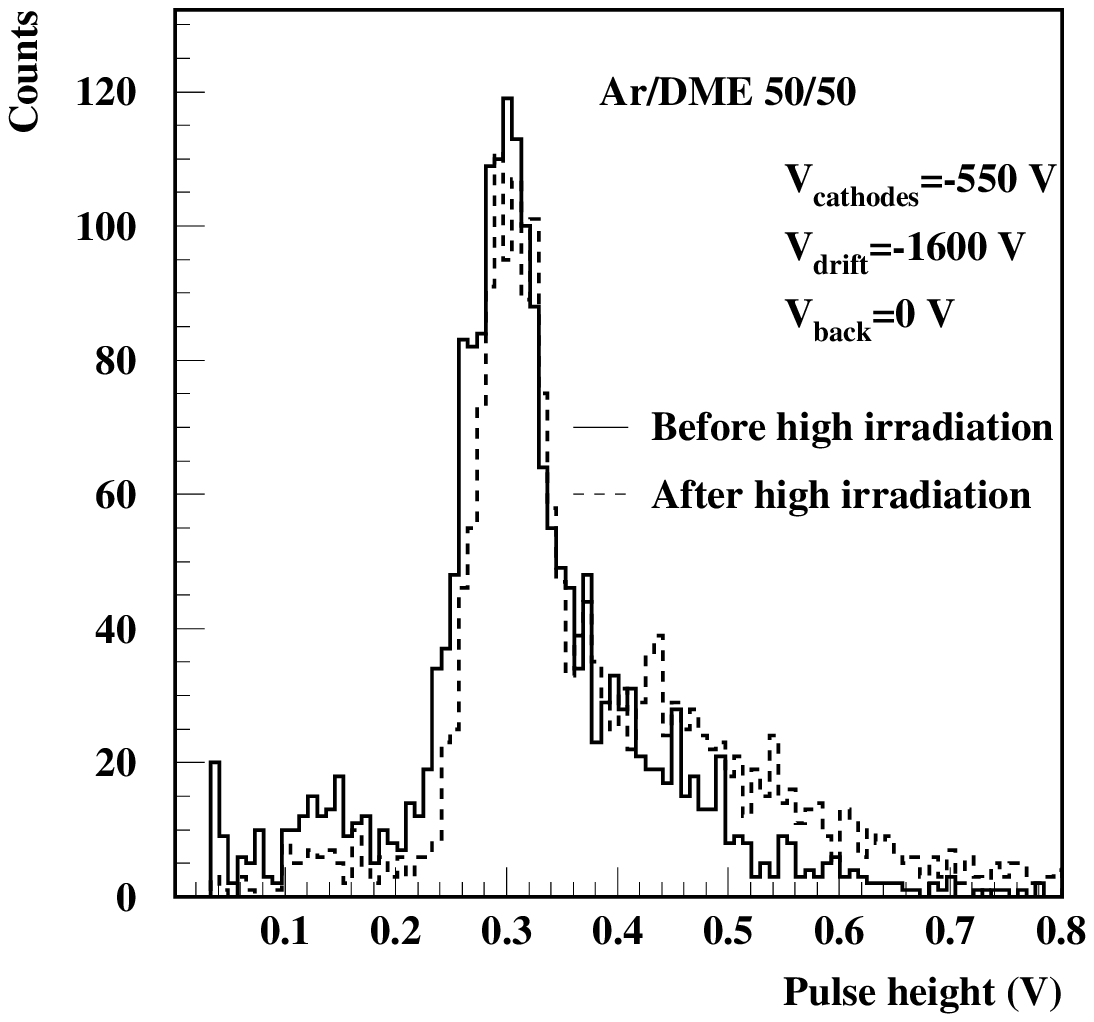,height=11cm}}

\vspace{.5cm}

Figure 9

\mbox{\epsfig{file=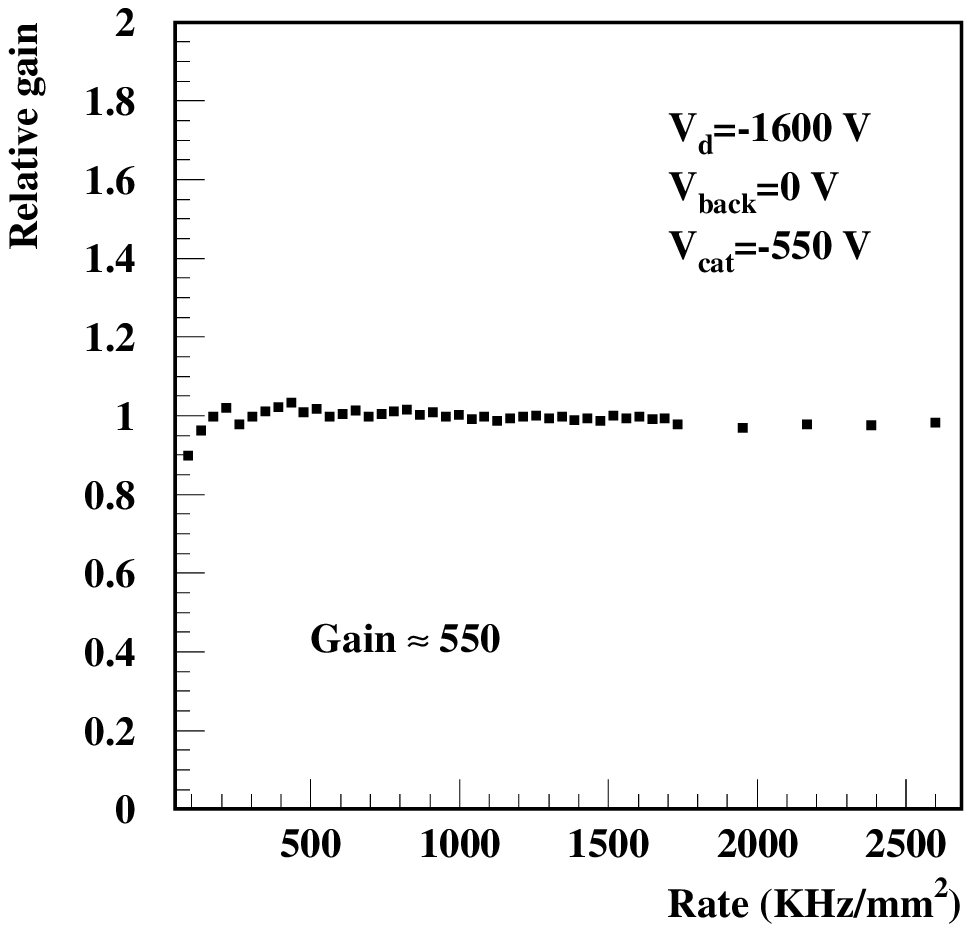,height=13cm}}

\vspace{.5cm}

Figure 10
\end{center}
\end{figure}

\end{document}